\documentclass[preprint,showpacs,preprintnumbers,amsmath]{revtex4}
\usepackage{amssymb}

\usepackage{graphicx}
\usepackage{dcolumn}
\usepackage{bm}
\usepackage[latin1]{inputenc}

\begin{document}

\title{Doppler cooling to the recoil limit using sharp atomic transitions}
\author{V{\'e}ronique Zehnl{\'e} and Jean Claude Garreau}
\affiliation{Laboratoire de Physique des Lasers, Atomes et
  Mol{\'e}cules, Centre d'Etudes et de Recherches Laser et Applications,
Universit\'{e} des Sciences et Technologies de Lille, F-59655
Villeneuve d'Ascq Cedex, France}
\homepage{http://www.phlam.univ-lille1.fr}

\preprint{v 1.08}

\date{\today}

\begin{abstract}
In this paper, we develop an analytical approach to Doppler cooling
of atoms by one- or
two-photon transitions when the natural width of the excited level is so
small that the process leads to a Doppler temperature comparable to
the recoil temperature. A ``quenching'' of the sharp line is introduced in
order to allow control of the time scale of the problem. In such
limit, the usual Fokker-Planck equation does not correctly describe
the cooling process. We propose a generalization of the Fokker-Planck
equation and derive a new model which is able to reproduce correctly
the numerical results, up to the recoil limit. Two  cases of practical
interest, one-photon Doppler cooling of strontium and two-photon
Doppler cooling of hydrogen are considered.
\end{abstract}

\pacs{42.50.Vk, 32.80.Pj}

\maketitle
%

\section{Introduction}
\label{sec:Intro}
Laser cooling is one of the greatest achievements of
Atomic Physics in the few last decades, having open a thorough new field of
applications and led to the first experimental demonstration of the
Bose-Einstein condensation in dilute gases \cite{ref:FirstBEC}.
The simplest laser
cooling process is the well-known Doppler cooling by a red-detuned
standing wave
interacting with a two-level atom \cite{ref:Houches}. A usual way of
describing Doppler cooling is to derive a Fokker-Planck equation, leading to
a limit ``Doppler temperature" given by $T_D \equiv \hbar \Gamma/
(2k_B)$, where $\Gamma$ is the width of the transition and
$k_B$ is the Boltzmann's
constant. The Fokker-Planck equation is valid as long as the mean
velocity of the
atoms is large compared to the atomic velocity shift due to a
fluorescence cycle, which is of the order of the recoil velocity
$v_r = \hbar k_0/M$, where $k_0 = \omega_0/c$
is the wavenumber of the atomic transition and $M$ is the
mass of the atom. The
Doppler temperature should thus be much larger than the recoil temperature
$T_r \equiv M v_r^2/k_B$, which implies 
\begin{equation}
  \Gamma \gg k_0 v_r \text{ .}
  \label{eq:FPvalidity}
\end{equation}
(Note that $k_0 v_r$ is the Doppler effect associated to the recoil
velocity). In the present paper, we shall consider two cases in which the
inequality~(\ref{eq:FPvalidity}) is \emph{not} satisfied, namely the
one-photon Doppler cooling of strontium and the two-photon Doppler cooling
of hydrogen.

Doppler cooling of strontium has recently been studied by experimental
groups, with aims varying from achieving high phase-space densities
\cite{ref:CoolSr} to the study of the coherent backward
scattering of light by a disordered cold-atom cloud \cite{ref:CohBackS}. The
idea is to take advantage of the very sharp ($\Gamma \approx 7.5$ kHz),
spin-forbidden, line $^1S_0-^3P_1$ whose theoretical Doppler
temperature is comparable to the recoil limit.

Laser cooling of hydrogen atoms, on the other hand, is still a goal to be
achieved. Laser cooling of \emph{trapped} hydrogen atoms was experimentally
demonstrated \cite{ref:CoolHTrapped}, but cooling of \emph{free} atomic
hydrogen is still a defy for experimentalists. The main difficulties are
that \textit{i}) hydrogen forms tightly bonded molecules that should be
dissociated before any experiment on atoms can be done,
and \textit{ii}) the optical
transitions starting from the fundamental state all fall in the
vacuum-ultraviolet (VUV), e.g. at 121 nm for the $1S-2P$ transition.
Moreover, although Bose-Einstein condensation of hydrogen was achieved \cite
{ref:BECH} \emph{without} use of such techniques, laser cooling of hydrogen
would certainly improve and simplify the procedure leading to
condensation, which can be of considerable practical relevance. Cooling of
atomic hydrogen by {\em pulsed} two-photon transitions was suggested in 1993 by
M. Allegrini and E. Arimondo \cite{ref:Maria_Ennio} using $\pi$ pulses on
the $1S-3S$ transition (wavelength of 200 nm for two-photon transitions).
Two-photon transitions present the advantage of dividing the laser frequency
by two, which means that the required radiation falls in more favorable
wavelength domains. The recent progress on the generation of CW, VUV, laser
radiation, in the context of metrological studies of the $1S-2S$ transition 
\cite{ref:1S2S}, allows one to realistically envisage the two-photon Doppler
cooling of hydrogen in the \emph{continuous wave} regime on the very sharp $%
1S-2S$ transition (the lifetime of the $2S$ state is 0.2 s), as we suggested
in a previous work \cite{ref:DopplerTwoPhoton}.

Laser cooling relies on the ability of atoms to perform
a great number of fluorescence cycles in which momentum is exchanged
with the radiation field in a short time (compared e.g. to
the collision time).
From this point of view, sharp lines are not suitable for cooling, because
the cooling time becomes very long. On the other hand, the Doppler
temperature is proportional to the linewidth of the excited level (this
result is also valid for two-photon transitions); from this point of view,
sharp lines are interesting for cooling. In order to conciliate these
antagonistic characteristics of sharp transitions, we proposed in
Ref. \cite{ref:DopplerTwoPhoton} to use the ``quenching" of the
$2S$ state of hydrogen to
control the cycling frequency of the process and we shown that this process
allowed to approach the recoil temperature where the Fokker-Planck
equation breaks down. In the present work we develop a refined analysis of
the Doppler process valid up to the recoil limit both for one- and
two- photon Doppler cooling.

\section{Doppler Cooling with one- and two-photon transitions}
\label{sec:OneTwoPhoton}

\begin{figure}[tbp]
\includegraphics{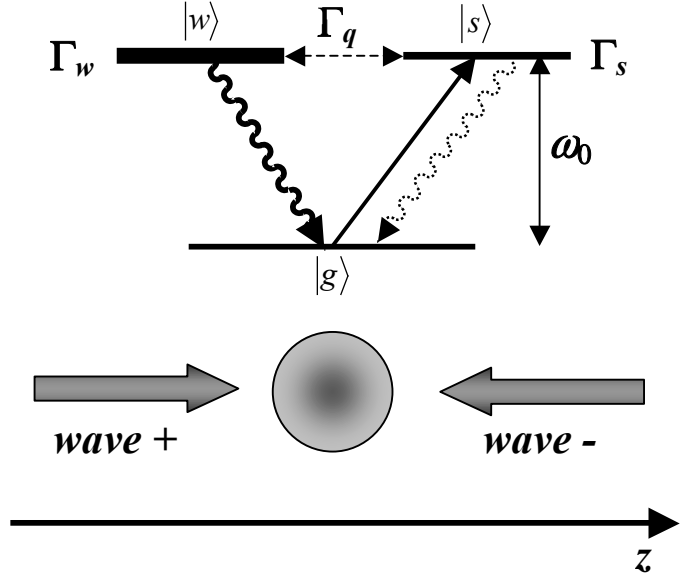}
\caption{Atomic levels involved in the cooling process. $|s\rangle$ is the
sharp level and $|w\rangle$ is the wide level. These levels are coupled by a
quenching process of rate $\Gamma_q$. Spontaneous emission from the
level $|s \rangle$ is very small.} 
\label{fig:LevelScheme}
\end{figure}

Let us consider an atom of mass $M$ and velocity $v$ parallel to the $z$%
-axis (Fig. \ref{fig:LevelScheme}) interacting with two counterpropagating
waves of angular frequency $\omega _{L}$ with $p\omega _{L}=\omega
_{0}+\delta $, ($p=1,2$ for resp. one- and two-photon transitions), $\omega
_{0}/2\pi =2.5\times 10^{15}$ Hz for H and $4.4\times 10^{14}$ for Sr). We
have also $k_0 \equiv \omega _{0}/c = pk_{L}$. We consider three
atomic levels, namely the ground state $|g\rangle $,
which is coupled by the radiation, via one- or two-photon transitions,
to a sharp level $|s\rangle $
of linewidth $\Gamma _{s}$, and the wide level $|w\rangle $ which is coupled
to the sharp level $|s\rangle $ through a quenching process of rate $%
\Gamma _{q}$ \cite{note:quenching} and to the ground-level $|g\rangle $ by
spontaneous emission (natural linewidth $\Gamma _{w}$). We introduce two
simplifying assumptions: \textit{i}) The wide level $|w\rangle $ can be
eliminated adiabatically, so that everything happens as if the sharp level
has, due to the quenching process, an effective linewidth 
\begin{equation}
\Gamma ^{\prime }=\Gamma _{w}\frac{\Gamma _{q}}{\Gamma _{q}+\Gamma _{w}}%
\equiv g\Gamma _{w}\text{ .}  \label{eq:EffGamma}
\end{equation}
where $g$ is a controllable quenching ratio.
\textit{ii}) We consider that the sharp and wide levels correspond to the
same transition frequency with respect to the ground state. This is a very good
approximation for the case of hydrogen, where the quenching couples levels $%
2S$ and $2P$, that are quasi-degenerated. In the case of the strontium
the transition frequencies differ by about 30 \%, but this does not
affect {\em qualitatively} our conclusions.
The velocity shift corresponding to spontaneous emission of a photon is thus 
$v_{r}\equiv \hbar k_0/M\approx 3.1$ m/s, or $T_{r}\approx 1.2$ mK for
H, and $v_{r}\approx 7$ mm/s and $T_{r}\approx 0.53$ $\mu$K, for Sr.

We shall now introduce some specifics of the two-photon Doppler cooling.
Four two-photon absorption process are allowed: \textit{i})
absorption of two photons from the $+z$-propagating wave (named wave ``$+$%
'' in what follows), with a rate $\Gamma _{1}$ and corresponding to the a
total atomic velocity shift of $+v_r$; \textit{ii}) absorption of two
photons from the $-z$-propagating wave (wave ``$-$''), with a rate $\Gamma
_{-1}$ and atomic velocity shift of $-v_r$; \textit{iii}) the absorption
of a photon in the wave ``$+$'' followed by the absorption of a photon in
the wave ``$-$'', with no velocity shift and \textit{iv}) the absorption of
a photon in the wave ``$-$'' followed by the absorption of a photon in the
wave ``$+$'', with no velocity shift. The two latter process are
indistinguishable, and the only relevant transition rate is that obtained by
squaring the sum of the \emph{amplitudes} of these process (called $\Gamma
_{0}$). Also, these process are ``Doppler-free'' (DF) as they are
insensitive to the atomic velocity (to the first order in $v/c$) and do not
shift the atomic velocity. Thus, they cannot contribute to the cooling
process. As atoms excited by the DF process must spontaneously
decay to the ground state, this process \emph{heats} the atoms. In the limit
of low velocities, the transition amplitude for each of the four processes
is the same. One thus expects the DF transitions to increase the equilibrium
temperature by a factor of two, which is verified both numerically and
analytically \cite{ref:DopplerTwoPhoton}. Note that the analysis of
the one-photon process can be deduced straightforwardly from the results
below by setting the Doppler-free term to zero. The transition rates
are given by \cite{ref:TwoPhoton}:
\begin{subequations}                                     
\begin{eqnarray}
  \Gamma _{\pm 1}(V) &=& \Gamma _{w} \frac{g}{2} \text{ }
  \frac {\bar{I}^{p}} {(\delta \mp KV)^{2}+g^{2}/4}\\  
\Gamma _{0}&=& 2\Gamma _{w} g \text{ }
\frac {(p-1)\bar{I}^{p}} { \delta^{2}+g^{2}/4 }
\label{eq:twophotontransrates}
\end{eqnarray}
\end{subequations}
where $p=1,2$ corresponds to, resp., one- and two-photon transitions.
$\Gamma_{\pm 1}$ describe, respectively, the absorption from the
``$+$'' and ``$-$'' waves, and $\Gamma_0$ the 
DF transition rate (note that
this terms is equal to zero for one-photon transitions). $\bar{I}\equiv
I/I_{s}$ where $I_{s}$ is the saturation intensity, $\delta$ is the
detuning divided by $\Gamma _{w}$, $K \equiv k_0 v_r /\Gamma _{w}\approx 0.26$
for H and $K\approx 3.2\times 10^{-4}$ for Sr, and $V\equiv v/v_{r}$.
The only qualitative difference between one- and two-photon Doppler
cooling, in the present context, is the presence of Doppler-free terms
in the latter case. Note that, with the above normalizations, the
usual Fokker-Planck approach is expected to break down when $g \leq
|\delta| \ll K$.

\section{Derivation of the generalized Fokker-Planck equation}
\label{sec:GFPE}

The rate equations describing the evolution of the velocity distribution $%
n(V,t)$ and $n^{*}(V,t)$ for, respectively, atoms in the ground and in the
excited level are 
\begin{subequations}
\begin{equation}
{\frac{\partial n(V,t)}{\partial t}}=-\big[ \Gamma _{-1}(V)+\Gamma
_{0}+\Gamma _{1}(V)\big] n(V,t)+{\frac{\Gamma^\prime}{2}}\big[
n^{*}(V-1)+n^{*}(V+1)\big]
\label{eq:rate_n}
\end{equation}
\begin{equation}
{\frac{\partial n^{*}(V,t)}{\partial t}}=\Gamma _{-1}(V-1)n(V-1,t)+\Gamma
_{0}n(V,t)+\Gamma _{1}(V+1)n(V+1,t)-\Gamma^\prime n^{*}(V,t)\text{ .}
\label{eq:rate_nstar}
\end{equation}
\end{subequations}
The deduction of the above equations is quite straightforward (cf. Fig \ref
{fig:LevelScheme}). The first term in the right-hand side of Eq.~(\ref
{eq:rate_n}) describes the de-population of the ground-state velocity class $%
V $ by two-photon transitions, whereas the second term describes the
re-population of the same velocity class by spontaneous decay from the
excited level. In the same way, the three first terms in the right-hand side
of Eq.~(\ref{eq:rate_nstar}) describe the re-population of the excited state
velocity class $V$ by two-photon transition, and the last term the
de-population of this velocity class by spontaneous transitions. For each
term, we took into account the velocity shift ($V \rightarrow V \pm 1$)
associated with each transition and supposed that spontaneous emission is
symmetric under spatial inversion.

For laser intensities below the saturation intensity of the
transitions, one can adiabatically eliminate the
population of excited level and reduce the Eqs.~(\ref{eq:rate_n})
and(\ref{eq:rate_nstar}) to one equation describing the evolution
of the ground-state population:
\begin{eqnarray}
{\frac{dn(V,t)}{dt}} &=&-\left[ \Gamma _{0}+\frac{\Gamma _{-1}(V)}{2}+ {%
\frac{\Gamma _{1}(V)}{2}}\right] n(V,t)+  \nonumber \\
&&{\frac{1}{2}} \Big\{ \Gamma _{0}\left[ n(V-1,t)+n(V+1,t)\right]
+ \Gamma_{-1}(V-2)n(V-2,t)  \nonumber \\
&+& \Gamma _{1}(V+2)n(V+2,t) \Big\}
\label{eq:elim}
\end{eqnarray}

The set of ordinary differential equations in
Eqs.~(\ref{eq:rate_n}) and (\ref{eq:rate_nstar}) or Eq.~(\ref{eq:elim})
has no known analytical solution and must be
solved numerically. However, it is possible to develop some analytical
approaches following the general idea leading to the Fokker-Planck
equation (FPE) for the time-evolution of the velocity distribution
(see Refs. \cite {ref:DopplerTwoPhoton}, \cite{ref:FPE}).
Taking the continuous limit of Eq.~(\ref{eq:elim}) with respect to $V$,
we obtain the following ``generalized Fokker-Planck equation''
(GFPE) \cite{note:continuous}: 
\begin{equation}
{\frac{\partial n}{\partial t}=}{\frac{\partial (f_{1}n)}{\partial V}}+{%
\frac{\partial ^{2}\left( f_{2}n\right) }{\partial V^{2}}}  \label{eq:FPE}
\end{equation}
where the functions $f_{1}(V)$ and $f_{2}(V)$ are given by: 
\begin{subequations}
\begin{eqnarray}
f_{1}(V) &=&\Gamma _{-1}(V)-\Gamma _{1}(V) \\
f_{2}(V) &=&\Gamma _{-1}(V)+\Gamma _{1}(V)+\frac{\Gamma _{0}}{2}\text{ .}
\label{eq:f1f2}
\end{eqnarray}
\end{subequations}
In this form, we keep explicitly the $V$-dependent coefficients $f_{1}(V)$
and $f_{2}(V)$. Note however, that this limit implies a
smooth variation of the velocity distribution $n(V)$ as well as of
the rates 
$\Gamma _{\pm 1 }(V)$ [on a $V$-interval which is $O(1)$]. However, the
rates $\Gamma _{\pm 1}(V)$ present sharp peaks at the values $V=\pm \delta /K$
(in the limit $g\rightarrow 0)$ and the GFPE is thus not valid
around these values. Note that the standard FPE can be obtained
from the above equations in the limit $V\rightarrow 0$: 
\begin{eqnarray}
f_{1}(V) &=&2V\frac{d\Gamma _{-1}(V)}{dV} \\
f_{2}(V) &=&2\Gamma _{-1}(V=0)+\frac{\Gamma _0}{2}
\label{eq:f1f2FP}
\end{eqnarray}
where the derivative is evaluated at $V=0$.

\begin{figure}[tbp]
\includegraphics{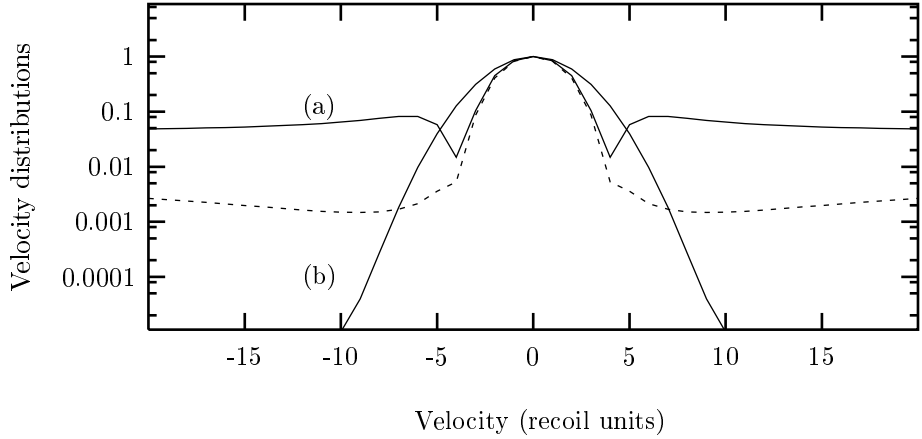}
\caption{Velocity distribution for two-photon Doppler cooling
  obtained from the GFPE [curve (a)], and FPE [curve (b)]
compared to the numerical simulations of Eq.~(\ref{eq:elim})
(dashed line). For the sake of comparison all distributions are set
to $n(V)=1$ for $V=0$. The distribution temperatures are
$T=2.19$ for (a),
$T^\prime = 3.88$ for (b) and $2.16$ for the numerical simulation
(temperature in recoil units). Parameters are $g=0.2$ and $K=0.26$. }
\label{fig:fv}
\end{figure}

To our knowledge, the above system has no
analytical solution. However, we can focus on
their asymptotic steady-state solution. By setting $\partial n{/}\partial
t\equiv 0$ in Eq.~(\ref{eq:FPE}), we straightforwardly obtain: 
\begin{equation}
  n(V)=\frac{1}{f_{2}(V)} \exp \Big\{ -\left[ \int
    {\frac{f_{1}(V)}{f_{2}(V)}dV} \right] \Big\}
  \label{eq:nformel}
\end{equation}
The steady-state distribution is then obtained by integration of Eq.
(\ref{eq:nformel})   ($f_1/f_2$ is a ratio of polynomials in $V$ and
the integration is obtained by standard methods). One gets:
\begin{equation}
  n(V)=\frac{1}{f_{2}(V)}\exp \left[ \frac{2\left( \delta
  ^{2}+g^{2}\right) \delta }{%
K\sqrt{\left( 7\delta ^{4}+22\delta ^{2}g^{2}-g^{4}\right) }}\allowbreak
\tan^{-1} \frac{2K^{2}V^{2}-\delta ^{2}+3g^{2}}{\sqrt{\left( 7\delta
      ^{4}+22\delta ^{2}g^{2}-g^{4}\right) }}\right]
\label{eq:GPFEDistr}
\end{equation}
[provided $\left( 7\delta ^{4}+22\delta ^{2}g^{2}/4-g^{4}/16\right) >0$ or
more simply $|\delta |\gtrsim g/10$. Note that this inequality is
fulfilled in most practical situations, since $|\delta|$ must be high
enough to allow the system to distinguish the ``+'' and the ``-''
cooling beams].
The solution of the FPE, on the other hand, is the gaussian
velocity distribution: 
\begin{equation}
  n^\prime(V)=\exp \left[ - \frac{1}{2}{\frac{|\delta| K}
      {\delta ^{2}+g^{2}/4}}V^{2} \right] \text{ .}
\label{eq:FPEDistr}
\end{equation}
These velocity distributions are shown in Fig.~\ref{fig:fv} and
compared to the results of a direct numerical simulations of
Eq.~(\ref{eq:elim}). Eq.~(\ref{eq:GPFEDistr}) compares
very well to the numerical solution for
$|V|<|\delta |/K\sim 4$ (that is, for the central peak of cold
atoms), while the  FPE distribution Eq.~(\ref{eq:FPEDistr}) leads to a
broader (hotter) distribution. Both approaches fail to
correctly describe the (uninteresting) background of hot atoms.

\begin{figure}[tbp]
\includegraphics{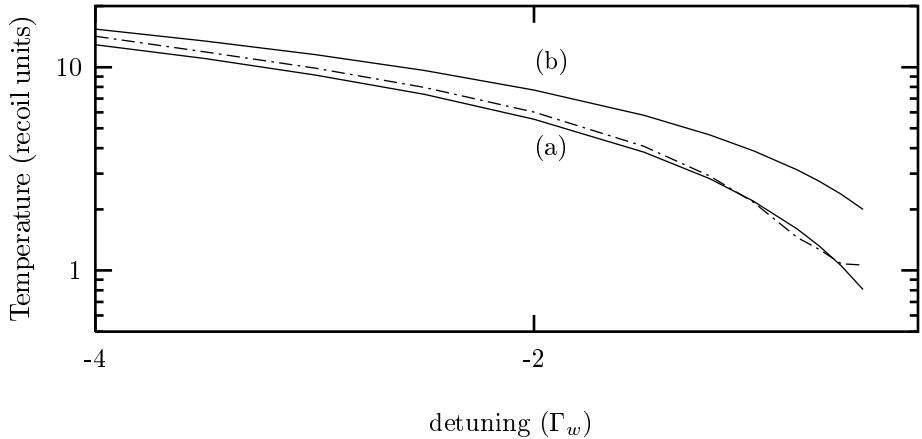}
\caption{Temperature vs. detuning for two-photon cooling. The curve (a)
corresponds to the GFPE and  the curve (b) to the FPE (both
represented by solid lines). The numerical values corresponds to the
dashed-dotted line). The agreement is clearly better for the GFPE.
Parameters are $g=0.2$ and $K=0.26$. }
\label{fig:temp}
\end{figure}

We can deduce the temperature of the distribution from its the central
peak of cold atoms (which typically corresponds to $|V|<|\delta |/K$,
that is, to atoms strongly interacting with the radiation) and we
neglect the wide background of hot atoms. In this limit, the
velocity distribution fits to a gaussian shape $\exp (-V^{2}/2T)$
where $T$ is the temperature in $T_{r}$ units.
For the GPFE, we obtain the following analytical expression: 
\begin{equation}
T=\frac{(\delta ^{2}+g^{2}/4)^{2}}{K\left[ -\delta ^{3}+3K\delta
    ^{2}-(\delta +K)g^{2}/4\right] }
\label{eq:TFPG}
\end{equation}
whereas for the FPE distribution the same approach gives: 
\begin{equation}
  T^\prime=\frac{\delta ^{2}+g^{2}/4}{K|\delta |}
  \label{eq:TFP}
\end{equation}
Fig.~\ref{fig:temp} compares both temperatures to the results of numerical
simulation. One observes that the values of $T^\prime$ increasingly
disagrees with the  numerical results as one approaches the recoil limit
(as expected), and predicts values that are typically too hot by a
factor $\simeq 3$ for detunings close the Doppler minimum $|\delta|
\simeq g/2$ [obtained from Eq.~(\ref{eq:TFP})]. For higher detunings
($|\delta| \gg K$),  both the FPE and
GFPE approach the same limit $T\simeq|\delta |/K$. Note that
for $|\delta| < g/2$  the GPFE approach also breaks down, leading to
sub-recoil temperatures. In this
case, the peak of cold atoms disappears into the background of hot
atoms, it is then hard define a temperature.
However, this regime is not of great experimental interest.

In the case of the cooling by one-photon transitions, we obtain the
velocity distribution simply by setting $\Gamma _{0}=0$
in Eqs.~(\ref{eq:f1f2}) and (\ref{eq:f1f2FP}):
\begin{eqnarray}
  n(V) &=&\allowbreak \frac{\left[ (\delta ^{2}+g^2/4)^2
  +K^{2}V^{2}(\delta^2+g^{2}/4)
\right] ^{\frac{\delta}{K} }}{f_{2}(V)|_{\Gamma _{0}=0}}  \nonumber \\
\allowbreak
n^\prime(V) &=&\allowbreak \exp \left(-\frac{|\delta| K}{\delta
  ^{2}+g^{2}/4}V^2 \right) \text{ .}
\label{eq:nSr}
\end{eqnarray}
for the GFPE and FPE cases respectively.
One deduces the temperatures of the cold central peak in the same way
as above.
The result is exactly $T/2$ and $T^\prime/2$. The factor two is a
consequence of the Doppler-free transitions that increases
the equilibrium temperature, as we pointed out in Sec.~\ref{sec:Intro}
and discussed in more detail in Ref. \cite{ref:DopplerTwoPhoton}.

\begin{figure}[tbp]
\includegraphics{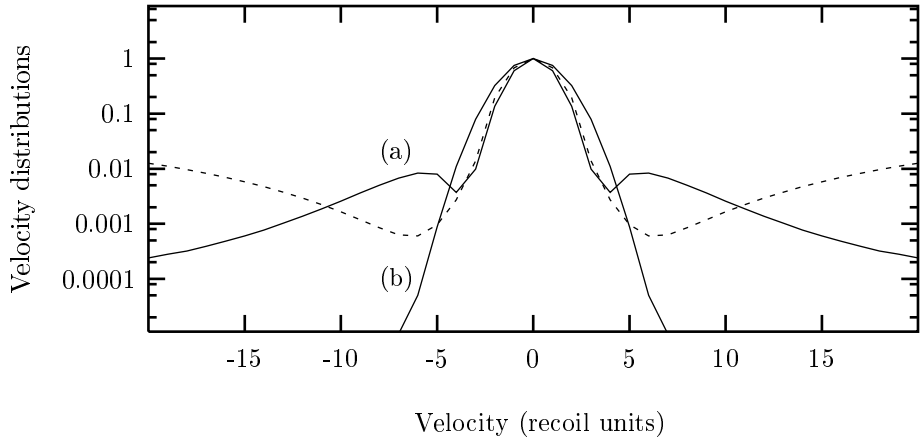}
\caption{Velocity distribution for one-photon transitions : GFPE
  [curve (a)], and FPE [curve (b)] (solid lines) and
  numerical simulation based on Eq.~(\ref{eq:elim}) (dashed line).
  The temperatures are $T=1.30$ for (a), $T^\prime=2.12$
  for (b) and $1.4$ for the numerical simulation. Parameters are
  $K=3 \times 10^{-4}$ for strontium,  $\delta = 4K$ and $g =0.8 K$.}
\label{fig:fvSr}
\end{figure}

Figure \ref{fig:fvSr} shows the distributions compared to the results
of a numerical integration of Eq.~(\ref{eq:elim}). The GFPE is seen to
correctly describe the velocity distribution and the temperature up to the
recoil limit. Note that
$\delta$ and $g$ must be of the order of $K$ in order to obtain
temperatures of the order of the recoil
temperature [see Eqs.~(\ref{eq:TFPG}) and (\ref{eq:TFP})].

In the case of hydrogen, a stray effect  can show up for the high
laser intensities: the photoionization of the excited state, causing a
decreasing in the number of cooled atoms. We evaluated the impact of
this effect in Ref. \cite{ref:DopplerTwoPhoton}, and showed that it
does not limit  the method.

\section{Conclusion}

In conclusion, we have developed a generalized Fokker-Planck approach
allowing to describe Doppler cooling on very sharp lines up to the
recoil limit, and showed that the resulting expressions compare very
well to the ``exact'' numerical. This approach can in principle
be extended to a more precise description of particular cases.

\begin{acknowledgments}
Laboratoire de Physique des
Lasers, Atomes et Molécules (PhLAM) is Unité Mixte de Recherche
UMR 8523 du CNRS et de l'Université des Sciences et Technologies de
Lille. Centre d'Etudes et de Recherches Laser et Applications (CERLA)
is supported by Ministère de la Recherche, Région Nord-Pas de Calais
and Fonds Européen de Développement Economique des Régions (FEDER).
\end{acknowledgments}


\begin{thebibliography}{99}
\bibitem{note:continuous} The functions $n$, $\Gamma_{\pm 1}$ depending
  on the discrete set of velocities $V,V \pm 1,V\pm 2$ are assumed to
  vary smoothly; the ``continuous limit''
  consists in replacing, in Eq.~\protect (\ref{eq:elim}),
 the function $n(V-1,t)$ (for instance) by: 
  $n(V-1,t)=n(V,t)-\partial_V n(V,t)-\partial^2_V n(V,t)/2 ...$,
and so on.
\end{thebibliography}

\end{document}